# APICURON: a reactive infrastructure for credit attribution across distributed research data ecosystems

Adel Bouhraoua[1], Mehdi Zoubiri[1], Gavin Farrell[1], Maria Cristina Aspromonte[1], Alex Bateman[2], Henning Hermjakob[2], Maria Victoria Nugnes[1], Daniela Raciti[3], Nicholas Stiffler[3], Geert van Geest[4], Ulrike Wittig[5], Karen Yook[3], Federica Quaglia[1,*], Silvio C.E. Tosatto[1,6,*]

***Co-corresponding authors**: [1]Department of Biomedical Sciences, University of Padova, Padova, 35121, Italy. **E-mails**: federica.quaglia@unipd.it, silvio.tosatto@unipd.it

[1] Department of Biomedical Sciences, University of Padova, Via Ugo Bassi 58/B, 35131 Padova, Italy
[2] European Molecular Biology Laboratory, European Bioinformatics Institute (EMBL-EBI), Wellcome Genome Campus, Hinxton CB10 1SD, United Kingdom
[3] Division of Biology and Biological Engineering, California Institute of Technology, Pasadena, California, 91106, United States
[4] Swiss Institute of Bioinformatics, Amphipôle, Quartier UNIL-Sorge, 1015 Lausanne, Switzerland
[5] Heidelberg Institute for Theoretical Studies, Schloss-Wolfsbrunnenweg 35, 69118 Heidelberg, Germany
[6] Institute of Biomembranes, Bioenergetics and Molecular Biotechnologies, National Research Council (CNR-IBIOM), Bari 70126, Italy

## Abstract

Data-driven biology relies on structured knowledge generated by expert biocurators, yet this work remains largely unrecognized in traditional academic assessments. To bridge this gap, we present the updated APICURON platform, a credit-attribution infrastructure that formally acknowledges these scientific contributions. Rather than relying on delayed batch reporting, the system captures curation events as they happen and transforms them into verifiable units of work. This design allows independent resources to define and update their own recognition models while preserving the historical record of each contribution. For researchers, APICURON highlights recent activity alongside lifetime achievements and connects verified activities to persistent academic profiles via ORCID. APICURON has been successfully integrated across biological knowledgebases and data resources, demonstrating its application to diverse workflows. Extending beyond biodata resources, it also supports recognition of non-traditional research artefacts, including training materials and research software, without imposing a rigid definition of contribution.

# Introduction

Modern scientific research increasingly depends on curated knowledgebases[1,2]. Databases of biological sequences, pathways, interactions, molecular complexes, genetic and phenotypic data, and computational models form the backbone of data-driven research, enabling reproducibility, interoperability, and large-scale analysis. In this context, biocuration[3], i.e., the intellectual effort required to extract, validate, and structure this knowledge, is essential to the integrity and quality of the scientific record. However, unlike publications or grants, these contributions lack systematic and timely mechanisms for formal attribution.

This attribution gap threatens the sustainability of community-driven scientific knowledge resources. We previously introduced APICURON[4], a platform designed to collect and recognize biocuration activities across partner knowledgebases. Since its first integrations in 2021, APICURON has recorded nearly 156,000 contribution events from 271 registered contributors across 18 partner resources as of May 2026, spanning biodata resources and training-related activities. The original implementation provided a framework for defining activity types, scoring schemes, badges, and leaderboards. However, contributions were processed at scheduled intervals rather than in real time as they occurred. This introduced delays between a curator's activity and its formal recognition, limited the ability to update scoring criteria rapidly, and made the recalculation of awards increasingly complex as the system grew. The present work expands the original APICURON implementation in both technical design and scope. Whereas the previous system demonstrated that biocuration activities could be collected and recognized across partner resources, here we address the requirements that arise when credit attribution becomes continuous, distributed and dependent on resource-specific recognition models. These include event-driven ingestion of contribution events, dynamic recalculation of credit under evolving recognition rules, temporal segmentation of indicators, grouping of contributions by contributor role, and recovery from interrupted processing.

To remain effective, scientific contribution tracking systems require the capacity to manage continuous streams of activity and evolving scoring rules. New contributions are added daily, and the way different activities are recognized may change over time. Systems that rely on periodic recomputation can become inefficient when even small adjustments to scoring rules require reprocessing large volumes of historical data, affecting both the availability and reliability of the service.

Credit attribution across research resources is not just a reporting problem. Unlike publications, many research contributions occur as small, repeated actions across databases, software platforms, training resources, and other shared infrastructures. These activities are often continuous, technically specialized, and defined differently by each research community. A useful recognition system, therefore, cannot rely on a single universal metric. It must capture contributions as they occur, preserve the local context that gives each activity meaning, allow recognition rules to evolve over time, and prevent comparisons between different research contexts. APICURON addresses this challenge by providing a shared infrastructure in which contribution events, recognition models and credit outputs remain distinct. This allows independent resources to define their own criteria for recognition while using a common mechanism for attribution, recalculation and dissemination.

Here, we present an updated APICURON as an extensible infrastructure for scientific recognition, redesigned around the principle that attribution should reflect the distributed nature of the resources it serves. By adopting a reactive design, the system treats both contribution events and modifications to resource-specific scoring models as unified triggers, enabling dynamic recalculation of the defined metrics and rewards as resources evolve while preserving global consistency. This architecture also allows rankings to be calculated efficiently across different time windows, such as weekly and monthly intervals, providing visibility into recent contributions alongside cumulative activity. Once verified, these activities can be linked to persistent researcher identifiers, such as the Open Researcher and Contributor ID (ORCID)[5], thereby transforming transient curation tasks into durable, discoverable components of the researcher's academic record.

APICURON's deployment across biological knowledgebases and emerging extensions to training and software-related activities demonstrate that the approach is not limited to a single database or contribution type, but provides a generalizable framework for recognizing research work that is poorly captured by publication-centric assessment systems.

# Results

## An event-driven architecture for credit attribution across research resources

The APICURON redesign was guided by four requirements: (i) contribution records should remain separate from the rules used to evaluate them, (ii) credit should be recalculable when recognition models change, (iii) indicators should remain scoped to the context in which work was performed, and (iv) recognition should be timely enough to support contributor feedback rather than only retrospective reporting. The original implementation of APICURON provided a mechanism for collecting and reporting biocuration activity across partner resources, but its batch-processing model was optimized for consistency at operational scale rather than continuous recalculation. Here, we extend this approach by modelling credit attribution as a dynamic process in which both contribution records and changes to recognition criteria can trigger updates to downstream outputs. In this architecture, the record of an activity is separated from the rules used to score and reward it, allowing recognition to be recalculated as resource-specific standards and expectations evolve. This separation between stable contribution events and mutable recognition models is a key design feature that allows credit records to remain durable while scores, badges, medals, and leaderboards can evolve with standards adopted by research communities. Transitioning to a reactive architecture allows APICURON to process updates with higher granularity and to support increasing volumes of contribution data. This shift shortens the reward loop for contributors and provides the structural reliability needed to accommodate the increasing volume and diversity of contribution streams (Figure 1).

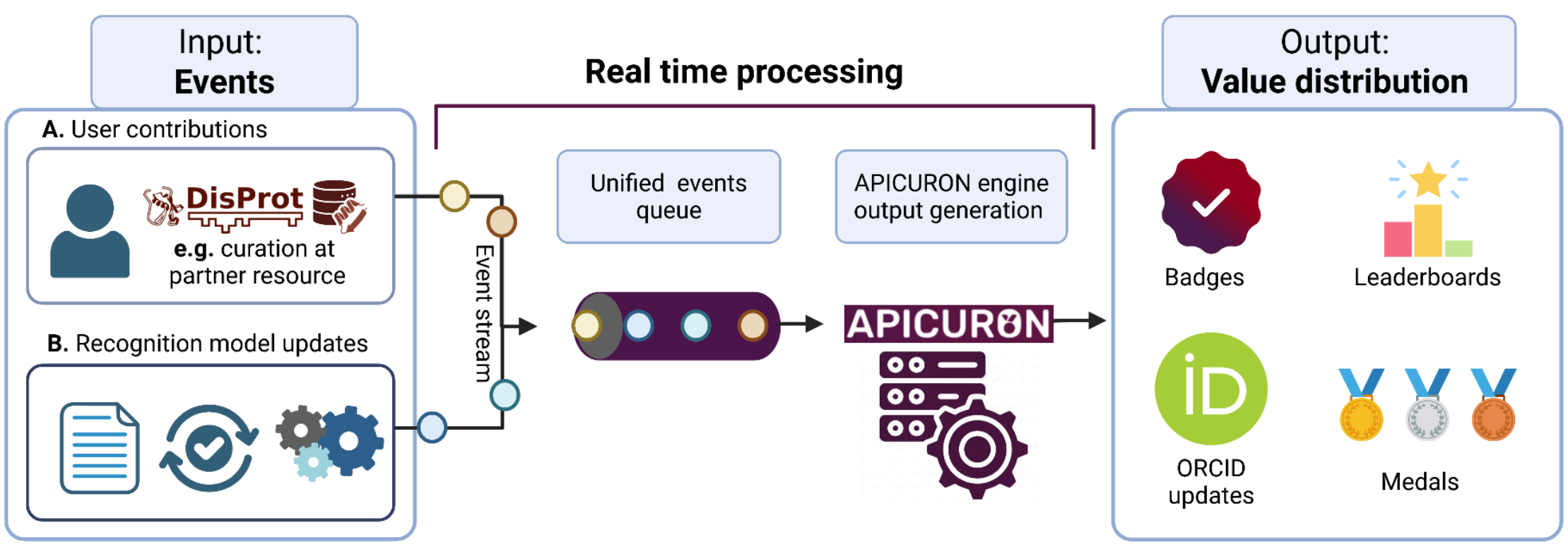

**Figure 1**. **APICURON event-driven processing architecture.** Inputs consisting of both continuous user contributions and discrete recognition model updates are channelled into a unified events queue. The core APICURON engine processes this queue asynchronously to update badges, leaderboards, medals, and ORCID synchronizations.

The outputs generated by APICURON depend on two independent states: the continuous stream of contribution events and the evolving definitions of recognition models. Changes to each introduce cascading updates to the curator profile, rewards, and leaderboards. To address these recurring computations, APICURON introduces an event-driven processing engine which allows non-blocking metric aggregation and reward updates.

## Resource-specific recognition models and rule updates

To accommodate changing scientific standards, the redesigned architecture enables recognition models to be updated without disrupting system stability. Each partner resource maintains autonomous control over its specific configuration (Figure 2), defining activity weights and reward criteria independently. Generated indicators are scoped to the intersection of a user and a partner resource, reflecting differences in recognition models across resources. One platform may emphasize manual annotation effort, whereas another may prioritize software development or training activity. This design does not attempt to reduce all contributions to a single cross-resource score. Instead, it provides a common attribution framework while allowing each resource to define credit in a way that reflects its own workflows, standards and community expectations.

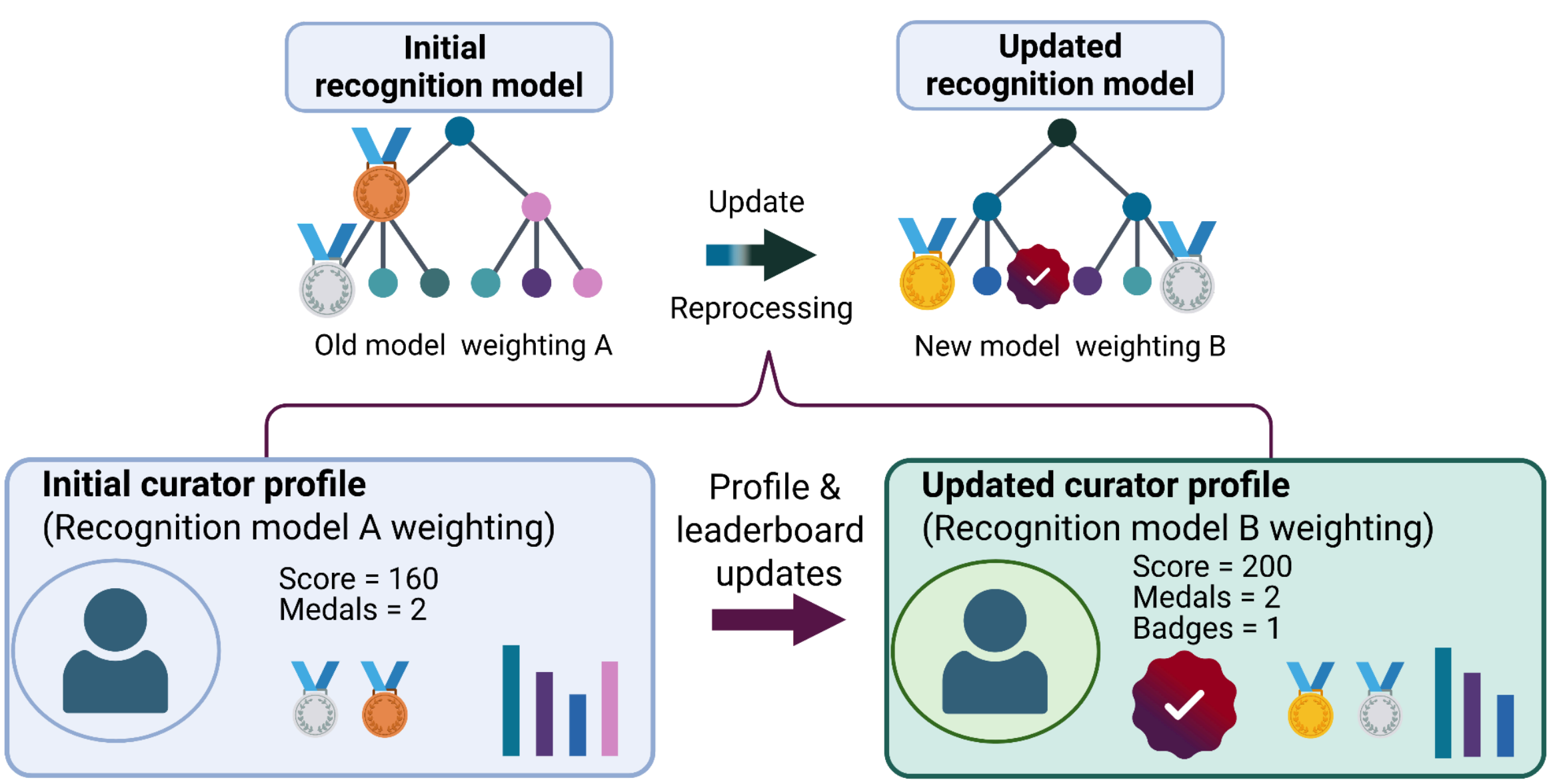

**Figure 2. Dynamic recognition model recalculation.** As a partner resource updates its specific recognition model, the system reactively re-evaluates contributions. The curator profile updates to reflect the new weighting scheme, adjusting scores and achievements as community standards evolve.

## Efficient recalculation and recovery under changing scoring rules

In a batch-oriented system, propagating changes to recognition rules and recalculating existing awards can be computationally intensive, limiting scalability. Modifications to recognition model criteria, such as changing the value of a contribution type or updating badge and medal definitions, often require recalculating previously awarded rewards to maintain consistency and data integrity. However, relying exclusively on this periodic approach forces the system to recalculate stable data to capture even minor changes, resulting in unnecessary computational overhead. Processing is orchestrated through an internal task queue that isolates event producers from consumers, enabling asynchronous execution while keeping event submission and event processing decoupled.

The system unifies stream and bulk processing by formalizing all state changes as discrete events. This unification also supports crash recovery and state reconciliation. In the event of service interruption, persisted raw reports are used to re-establish a consistent baseline. Upon restart, unprocessed records are identified and re-ingested through the standard bulk event pipeline, allowing the legacy bulk mechanism to function as a native recovery strategy without requiring external transaction managers.

Processing logic adapts dynamically to the type of event handled, invoking specific handlers for new contributions, badge definition updates, or medal recalculations. As a result, system complexity is managed within the recognition logic rather than at the infrastructure level. Together, these features establish APICURON as a generalizable credit-attribution infrastructure: one that preserves contribution records, supports evolving recognition models, separates recent activity from cumulative contribution, and maintains contextual fairness across different research communities.

## Temporal segmentation to enhance contributor visibility

Leaderboards in APICURON have been redesigned to support multiple time windows (Figure 3). Previously, each partner resource had a single leaderboard ranking curators by their total score, a weighted count of contributions defined by the resource. This design captured only the cumulative volume of contributions over the entire lifetime of the resource for each user, without filtering or weighting recent activity. While this approach accurately reflected absolute historical effort, it failed to weigh recent activity. Consequently, these systems can suffer from ranking stagnation, reflecting the Matthew effect[6] within digital curation, where the accumulated advantage of veteran contributors entrenched them permanently at the top. This can discourage newcomers from participating and make active, present-day curation efforts less visible. To overcome this barrier, APICURON's redesigned, event-driven architecture facilitates the simultaneous generation of time-bound indicators alongside continuous metrics. Consequently, the system now features three ranking views, comprising weekly, monthly, and cumulative leaderboards, which serves as a mechanism to broaden visibility, particularly for new contributors. By isolating recent activity into distinct temporal segments, the architecture increases immediate visibility for the momentum of new contributors, rewarding current engagement and consistency without changing historical, lifetime scores.

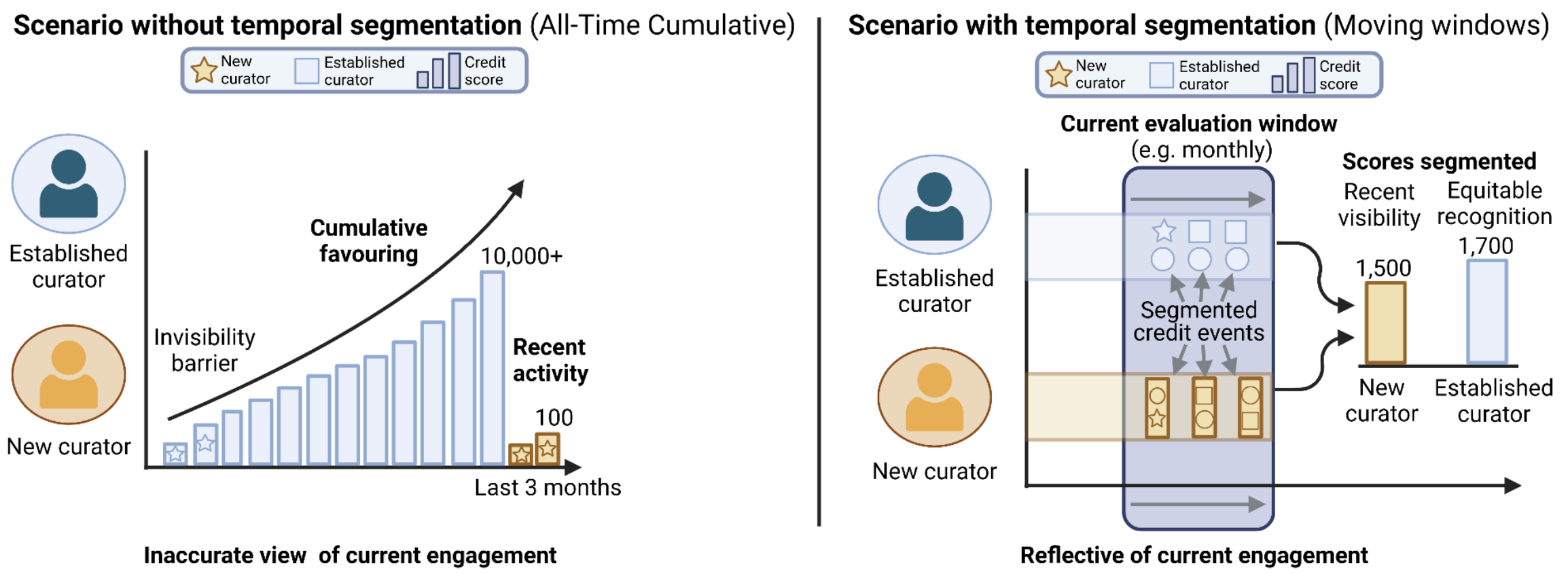

**Figure 3. Temporal segmentation of indicators.** Traditional cumulative ranking systems often obscure recent efforts by heavily favouring established curators. Segmenting credit events into discrete time windows enables both new and established curators to be evaluated fairly on their current activity, mitigating ranking stagnation and increasing visibility.

Recognizing that curation dynamics and community workflows vary across biological databases, this temporal segmentation is coupled with extensive flexibility. By default, the APICURON interface prioritizes the monthly leaderboard; this serves as a balanced baseline, acting as a middle ground between high-volatility weekly metrics and static cumulative rankings. However, resource managers are equipped with a dashboard allowing them to alter this default temporal view to suit the specific pace of their community. Furthermore, administrators can selectively configure which attribute columns are rendered on the leaderboard, ensuring that the public attribution interface remains tightly aligned with the specific goals and needs of each partner platform.

Temporal segmentation, therefore, functions not only as an interface feature but also as a metric-design strategy that separates lifetime contribution records from current engagement signals.

## Role-aware contribution grouping for fair comparison

A key challenge in designing a credit and recognition research infrastructure is the disparity between different types of contributors. By quantifying effort and publishing performance indicators and reward mechanisms, a system creates performance indicators that invite comparison. Comparing the high-volume output of professional, often-salaried biocurators with the sporadic contributions of community volunteers can misalign metrics, risking the undervaluation of meaningful volunteer work and the misattribution of professional output.

To mitigate this, APICURON introduces Contribution Groups (Figure 4), allowing biological databases and research platforms to define contributor cohorts and isolate metric aggregation. Each cohort can benefit from an adequate recognition model and associated reward mechanisms, ensuring that competition and ranking occur only among peers with comparable resources and time commitments.

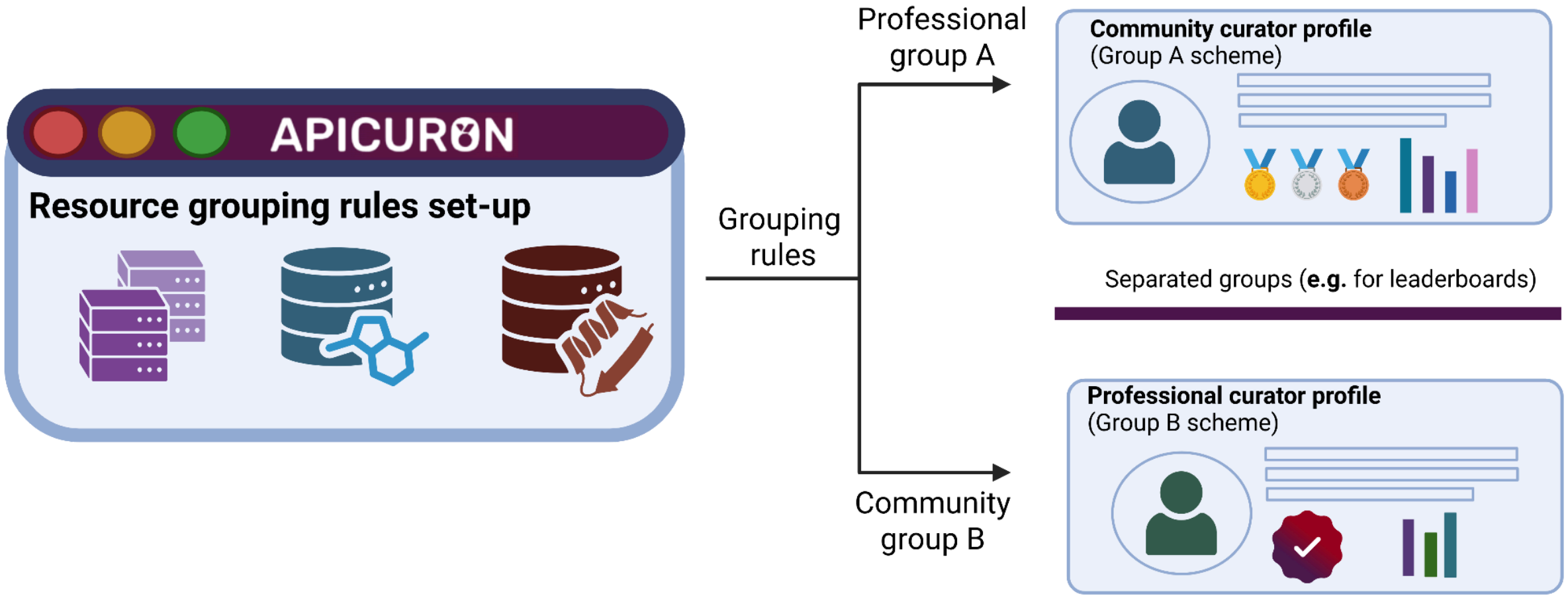

**Figure 4. Role-aware contribution grouping.** Resource grouping rules isolate metric aggregation based on distinct contributor roles. This separation allows platforms to establish tailored recognition schemes for community volunteers and professional staff, ensuring fair comparison and relevant rewards within appropriate cohorts.

This feature avoids the limitations of static user tagging. Users are not assigned only to a single global group; instead, APICURON utilizes event-level contextualization. Group membership is captured at the moment of the contribution, effectively tagging each activity report rather than the user profile. This provides a mechanism for preserving the provenance of the social and organizational context in which a contribution was made. This approach preserves the historical context of each contribution, particularly when researchers change roles. For example, if a community volunteer is later hired as core staff, contributions made before the transition remain associated with the volunteer group, while subsequent contributions are credited to the staff group. A single researcher can therefore accumulate distinct indicators and rewards for different phases of their career, without retroactively changing earlier records.

Requiring platforms to explicitly tag every API payload would impose an integration burden. To resolve this trade-off, APICURON implements a state-based defaulting mechanism, whereby administrators manage cohort assignments via the user interface by setting a user-group binding state that establishes the default destination for future contributions. If a report lacks a group tag, the ingestion engine resolves the user's current binding and assigns the configured default group, while still allowing explicit override on a per-report basis.

Modifying a user's default group binding does not retroactively alter previous data but redirects subsequent events to the new cohort. If a researcher transitions roles, for example, from community volunteer to professional staff, subsequent activity is credited to the new cohort while prior contributions remain immutably associated with the original group. By preserving this historical context through event-level tagging, the system prevents retroactive data distortion and enables a single researcher to accumulate distinct metrics for different phases of their career.

## Adoption across diverse biological resources

While APICURON was initially introduced as a framework piloted with three resources, i.e., DisProt, the Protein Ensemble Database (PED), and Biomappings, its subsequent evolution has been feedback-driven, with real-world implementation informing its architecture. To evaluate the platform's performance, interoperability, and community adoption across diverse curation environments, an implementation study was conducted in collaboration with an expanded cohort of biological knowledgebases. This group (Table 1) was selected to represent a broad range of activities involved in biocuration, from the granular manual annotation of intrinsic disorder to the automated classification of protein families. These integrations provide a practical test of whether a single credit-attribution infrastructure can support heterogeneous definitions of research work. Across the cohort, contribution events differ in granularity, review process, and domain semantics, yet they can be represented through the same event, recognition model, and reward abstractions. These early adopters were key participants of the international ELIXIR Europe Data Platform[7] and those who held global accreditations recognizing their importance in the biodata landscape, such as the ELIXIR Core Data Resources (CDR)[8], ELIXIR Deposition Databases (EDD) and Global Core Biodata Resource (GCBR)[9] status. Together, these resources represented a diverse international cohort with strong pan-European participation and broader international collaboration. This cohort was key to piloting APICURON and shape the systematic crediting of a broad range of biocuration activities within their respective resources and biological data types. Through successful initial APICURON use that met the needs of these resource stakeholders in real curation workflows, the platform was refined to accommodate the nuances and practical complexities of data annotation, quality control, tool development, and other essential contributions to biological data resources. This iterative deployment strengthened the system's flexibility and robustness, ensuring it could reliably capture and recognize heterogeneous forms of expert curation effort.

**Table 1. APICURON resource integrations and contributor metrics (2021–2026).** Event counts indicate the number of reports available in APICURON at the time of data extraction. Together, these integrations represented 18 partner resources and nearly 156,000 recorded contribution events as of May 2026, spanning biodata resources and training-related activities. Resources with zero events were onboarded or under configuration at the time of data extraction and had not yet submitted production contribution reports. All metrics represent a static snapshot as of May 2026. As APICURON is a live system, contributor counts and activity figures are subject to change.

| Resource | Description | Year Adopted | Curation events | Accreditation | Type |
| --- | --- | --- | --- | --- | --- |
| DisProt[10] | Database of intrinsically disordered proteins and regions with experimental evidence. | 2021 | 59337 | EDD | Biodata resource |
| Protein Ensemble Database (PED)[11] | Open access database for the deposition of structural ensembles, mainly intrinsically disordered proteins. | 2021 | 886 | ELIXIR Resource | Biodata resource |
| Biomappings[12] | Community-curated and predicted equivalences and related mappings between named biological entities. | 2021 | 7,753 | - | Biodata resource |
| Bioregistry[13] | Open source, community-curated registry, meta-registry, and compact identifier resolver. | 2022 | 0 | - | Biodata resource |
| Pfam[14] | Protein families database defined by sequence alignments and hidden Markov models. | 2022 | 36,159 | ELIXIR Resource | Biodata resource |
| PomBase[15] | Comprehensive model organism knowledge base for the fission yeast Schizosaccharomyces pombe. | 2023 | 0 | CDR, GCBR | Biodata resource |
| Rfam[16] | Resource for non-coding RNA families, secondary structures and covariance models. | 2023 | 5,725 | ELIXIR Resource | Biodata resource |

| BioModels[17] | Repository of curated mathematical models of biological and biomedical systems. | 2023 | 982 | EDD | Biodata resource |
|---|---|---|---|---|---|
| Reactome[18] | Peer-reviewed pathway database documenting human biological processes. | 2023 | 28,591 | CDR, GCBR | Biodata resource |
| IMEx/IntAct[19] | Molecular interaction database derived from literature and user submissions. | 2023 | 7,972 | CDR, GCBR | Biodata resource |
| Complex Portal[20] | Curated encyclopaedic resource of stable macromolecular complexes. | 2023 | 6,518 | - | Biodata resource |
| SABIO-RK[21] | Database of biochemical reaction kinetics, rate equations and experimental parameters. | 2024 | 958 | ELIXIR Resource | Biodata resource |
| DOME Registry[22] | A registry that allows users to contribute annotations, subject to rigorous manual review. | 2024 | 316 | ELIXIR Resource | Biodata resource |
| ELIXIR Training Materials[23] | Training materials developed by the ELIXIR Training Platform. | 2024 | 0 | ELIXIR Resource | Training |
| PDBe[24] | European resource for the collection, organization and dissemination of data on biological macromolecular structures. | 2024 | 78 | CDR, GCBR | Biodata resource |
| Glittr.org[25] | Curated database of educational material for the computational life sciences. | 2024 | 20 | ELIXIR Resource | Training |
| microPublication Biology[26] | Publishes brief, single-figure articles on research findings enabling rapid dissemination results. | 2025 | 607 | - | Biodata resource |
| S3 School[27] | Training program designed to teach good and modern coding practices tailored for scientific software. | 2026 | 33 | - | Training |

This initial phase of integration established APICURON as a framework for capturing credit and recognizing the intellectual effort underlying different biocuration practices. The following biological resource partners represent the first cohort of early adopters, demonstrating the system's applicability across distinct curation contexts. In each case, integration of APICURON contributed to improved visibility, transparency and consistency in recognizing biocuration activities, compared with previous in-house or ad hoc approaches.

Considering specific adoption cases, APICURON has captured increasing numbers of curated contributions as new resources adopted the platform (Figure 5). DisProt, the database of intrinsically disordered proteins, was the first biological resource integrated into APICURON, establishing a collaboration that has driven the co-evolution of both platforms. Because of the complexity of the DisProt biocuration workflow, APICURON evolved from a generic contribution-tracking system into a platform capable of representing every stage of the biocuration process in DisProt. Each contribution, from the generation of new protein entries or addition of experimental details according to the Minimum Information About Disorder Experiments (MIADE) guidelines[10], is captured and recognized. The system also mirrors collaborative effort behind each DisProt release, where curators are rewarded for their effort and participation in annotation rounds. Moreover, separate recognition pathways for volunteer and professional curators ensure a fair evaluation of contributions. To date, it has recorded more than 57,000 curation events from over 90 contributors. Similarly, for literature-driven resources such as IntAct, APICURON ensures the systematic crediting of continuous molecular interaction curation. By capturing actions that span the generation, revision, and invalidation of data, alongside publication approval processes, the infrastructure transforms routine editorial and quality control tasks into verifiable scientific contributions. Extending beyond traditional biological databases, APICURON has also been adopted by Glittr.org[25], a curated repository of educational material for the computational life sciences. This integration highlights the capacity of the system to capture and reward the curation of training materials, thereby acknowledging the key effort required to increase educational capacity within the research community. Collectively, these deployments demonstrate the adaptability of APICURON to heterogeneous workflows, ranging from structural annotation and interaction extraction to the curation of educational resources. The observed diversity of event types and reporting volumes has been a major driver for evolving the platform into a standardized credit infrastructure.

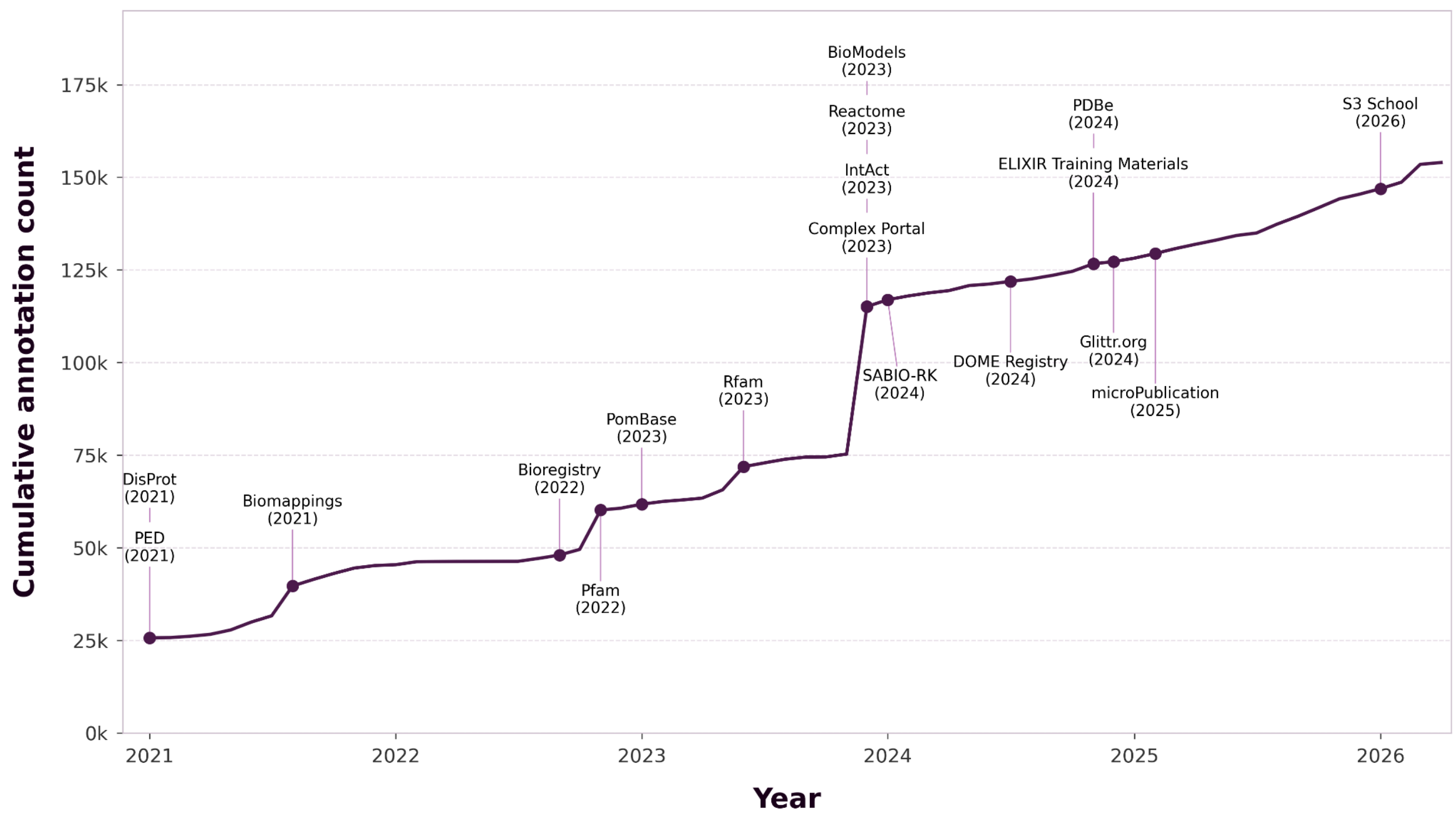

**Figure 5. Growth of curated contributions to APICURON over time (2021-2026), annotated with resource integration events.** Each annotation point marks the onboarding of a new partner resource into APICURON, with visible step increases in the cumulative report count following integration events. The steepest growth occurs around early 2024, coinciding with the simultaneous integration of several high-volume resources, including Complex Portal, IntAct, Reactome, and BioModels. Cumulative reports now account for nearly 156,000 recorded contribution events as of May 2026 across all integrated resources.

Following a successful pilot phase from 2022 to 2024, APICURON received renewed funding from ELIXIR Europe[7] to support further development and new service integrations. This funding has enabled extensive developer documentation, dedicated user support, and improved API guidance to maximize the platform's value. This improved support for uptake and integration after the pilot phase contributed to an important milestone, broader adoption and alignment across the international biocurator community. In 2025, APICURON was awarded formal recognition by the International Society of Biocuration[28] organization. This provides an official global endorsement for its uptake and integration into new resources, from which APICURON is seeing continued adoption across the life-sciences.

# Discussion

## Recognition as a component of research infrastructure

APICURON provides a framework to reduce the attribution gap that often leaves non-traditional research contributions unacknowledged. The system uses connections with partner resources to automate the tracking, acknowledgement and dissemination of diverse research activities, with biocuration serving as a primary use case, using identifiers such as ORCID to reliably attribute credit. APICURON shows how credit attribution can be embedded in the operation of research resources rather than added after contributions have occurred. By separating contribution events from recognition rules and credit outputs, the system supports common attribution workflows without requiring a universal definition of contribution.

Contributors generate activity records through their engagement with partner resources, which automatically transmit user contribution data to APICURON via API connections. Once ingested, the internal engine analyses these records to quantify activity by computing real-time statistics, updating global leaderboards, and assigning badges and medals to reward continuous engagement.

Following this analysis, credit is disseminated across multiple channels: the contributor's personal portfolio on the APICURON web interface is updated immediately, and validated records are pushed to third-party platforms, most notably synchronizing with ORCID so that these scholarly contributions become visible and citable alongside traditional research outputs. A customizable widget is also provided for resources to display the aggregated leaderboards on their website. All of this information is also available on request through the public Open API.

## Design strategies to prevent metric distortion

APICURON metrics are scoped to the intersection of a user and a partner resource, such that every measured indicator is defined by the unique combination of a researcher and a research platform. This scoping is critical because recognition models are heterogeneous across resources. Because one platform may prioritize manual effort while another weights scientific impact, cross-platform comparison is not meaningful by design. Researchers are assessed solely within the context of the specific recognition model they engaged with, as indicators across infrastructures are often distinct due to different activity granularities and weighting schemes.

By adding a temporal dimension, e.g., weekly or monthly intervals, to the user–resource pair, the system addresses ranking stagnation and enables platforms to highlight recent activity, mitigating inertia and increasing the visibility of recent contributors. Furthermore, to avoid the distortion that occurs when comparing the high-volume output of professional curators with the sporadic contributions of community volunteers, APICURON utilizes Contribution Groups. These allow research platforms to define logical cohorts and isolate metric aggregation, ensuring that competition and ranking occur only among peers with comparable resources and time commitments.

## Extending credit attribution beyond biocuration

While APICURON has proven to be an effective tool for recognizing biocuration efforts across multiple resources, its roadmap includes expansion to other non-traditional research artefacts beyond biodata resources. The ELIXIR STEERS 2025 Policy Brief on Credit and Recognition[29] and its related mapping of non-traditional research artefacts[30] highlight the need to better understand and recognize these outputs. These outputs highlight that non-traditional research artefacts require technical capacity to be captured for credit, particularly when they are produced by researchers and digital research technical professionals but overlooked in traditional research assessment frameworks[31]. Extending APICURON to manage these artefacts is a key step toward integrating diverse contributions into researcher assessment criteria, aiding global initiatives like the Coalition For Advancing Research Assessment (CoARA)[32] in supporting a healthier scientific research ecosystem.

Current work includes pilots to extend the credit capture to new categories, namely research software through the European Open Science Cloud (EOSC) EVERSE project[33] and training materials supported by ELIXIR Europe. Research software efforts focus on outputs such as tools and research infrastructure that deliver substantial value to the scientific community. Proof-of-concept technical pilots have already demonstrated the ability to capture granular development activities, including code contributions, software releases, and documentation. For training materials, successful integrations have been completed for the Glittr.org resource[25], supported by the Swiss Institute of Bioinformatics (SIB), and the S3 School training materials of EOSC EVERSE[27]. Additional support is being rolled out for community-hosted Git training materials using standard FAIR lesson templates by the ELIXIR Training Platform[7], with future plans for integrations with other training resources, such as the TeSS[34]. These efforts represent a step toward extending APICURON to new domains, with further extensions possible for peer review and other under-recognized research outputs.

Looking beyond APICURON as a standalone solution, the approach highlighted in the EOSC EVERSE Recognition Framework[35–37] addresses digital research outputs, including software, workflows, and training materials, that often remain undervalued. Despite being critical to scientific progress, these contributions frequently go unrecognized in publication-centric frameworks, stifling career progression for Digital Research Technical Professionals (dRTPs) who maintain the infrastructure of modern research. By adopting a federated architecture, APICURON can align with complementary platforms such as BIP! Scholar[38] and ORCID[5]. Together, these platforms provide the coverage required to capture granular research activities and make them visible in research assessment contexts. This integrated approach positions APICURON as a core component of the broader recognition ecosystem, supporting the formal attribution of non-traditional research artefacts and the specialists who develop them across diverse domains.

## Current constraints and implementation boundaries

APICURON's scope is bounded by what partner resources choose and are able to report: the system can only recognize contributions that are actually submitted, so the resulting indicators depend on the completeness and consistency of submitted activity records. Because each resource defines its own recognition model, the system is not intended to produce a single score for comparing researchers across platforms. The use of identifiable contribution records and ORCID-linked profiles also requires appropriate governance, a consent management system and privacy safeguards.

The initial implementation of APICURON operated as a periodic pipeline, introducing delays between contribution and recognition and making recalculation of awards computationally intensive when recognition model criteria changed. While the re-architecture into a reactive, event-driven system enables real-time processing of events and model updates, this shift introduces architectural complexity. Specifically, changes to recognition rules require recalculating previously awarded rewards to maintain data integrity.

The system also depends on coordinated governance and authentication structures. Every write operation must be traceable to a human actor; identity verification is delegated to ORCID, and access control is enforced through a Role-Based Access Control layer. While this ensures accountability, it requires consistent management of user roles and resource affiliations. Finally, although the unified event abstraction supports both real-time updates and bulk submissions, the system must accommodate large-scale re-ingestion and recovery workflows when required, reflecting the practical constraints of distributed research infrastructures.

## Future directions for scalable credit attribution

The re-architecture of APICURON into a reactive system enables real-time processing of granular contribution events and instantaneous response to modifications in recognition models. To support future scalability, event processing logic is decoupled from the persistence layer, allowing the event-handling subsystem to operate independently and, if needed, be moved to dedicated infrastructure, such as a message broker. The current internal task queue isolates event producers from consumers, providing an abstraction layer that enables future horizontal scaling without requiring core logic to be rewritten as the system outputs depend on both continuous contribution streams and evolving models. This design will also support crash recovery: persistent raw reports allowing unprocessed records to be re-ingested after service interruptions, ensuring durability without complex external transaction management.

Future considerations for APICURON include the rapid pace of change in scientific research methods driven by the inclusion of artificial intelligence (AI) tools such as large language models in scientific work. For example, in biological knowledgebases, AI-driven scalable curation approaches are reducing bottlenecks to aid the curation of relevant knowledge from literature. Literature triage systems are seeing success in automatically identifying relevant literature[39], but a major bottleneck remains the manual human effort required to extract and verify content. However, with the emergence of novel AI-assisted curation tools, some of these activities can be successfully automated while retaining human-in-the-loop review, as demonstrated by recent efforts of the Monarch Initiative[40]. Therefore, to maintain relevance in an era of AI-assisted curation workflows and human-orchestrated agentic systems, APICURON must evolve its architecture to account for this increased credit and recognition complexity of such research artefacts. Future AI-assisted curation workflows may increase the need to distinguish automated processing from human validation, oversight and accountability.

## Toward a unified framework for the recognition of research artefacts

While APICURON is established in tracking biocuration outputs, its architecture already encompasses diverse non-traditional research artefacts, aligning with CoARA priorities and European initiatives such as ELIXIR-STEERS and EOSC EVERSE. These frameworks advocate for formalizing credit across datasets, software, workflows, and training materials to reflect the reality of modern scholarship. By adopting a collaborative framework, APICURON helps connect individual activity and professional recognition, ensuring granular contributions are translated into verified records through integration with platforms like BIP! Scholar and ORCID. To drive this forward with international alignment, the newly established ELIXIR Europe Focus Group on Credit, Recognition and Usage serves as a key collaborative venue. Engaging 24 ELIXIR member states, this group facilitates the alignment of infrastructures like

APICURON to design novel credit-capture mechanisms for non-traditional research artefacts, underscoring APICURON's vital role in these broader European initiatives.

As science transitions toward AI-assisted and agentic systems, real-time attribution may become an important mechanism for linking automated outputs to verified human expertise. Ultimately, APICURON provides the technical architecture to transform scientific credit from a narrow, publication-centric model into a comprehensive record of the broad spectrum of research contributions.

# Conclusion

The APICURON platform provides a reactive infrastructure for formally recognizing research contributions beyond traditional publications, with adoption across life-science data resources. It translates individual biocuration activities and other technical research outputs into verifiable credit records that can be recognized by partner resources and linked to researcher profiles. Real-time attribution may become increasingly important for linking automated outputs to verified human expertise as AI systems become more common in research workflows. APICURON therefore provides a technical framework for supporting a transparent and sustainable record of research contributions across diverse research outputs and resources.

# Methods

## Indicators, Engagement and incentive mechanisms

APICURON's primary function is to aggregate contribution events into indicators of researchers' efforts around non-traditional research artefacts, such as biocuration. These outputs serve a dual purpose: providing platforms with feedback on community engagement and offering contributors a context for their own contributions. In APICURON, a recognition model describes how a partner resource translates reported activities into indicators, scores and rewards. Because recognition models are defined independently by each partner resource, scores are interpreted only within a given resource, temporal window and, where configured, contribution group. Building on its previous iteration, APICURON now includes features that offer increased flexibility in reward mechanisms, further incentivizing consistent contributions.

Partner resources define activities by assigning a "weight", a numerical value representing an activity's base value relative to others. This serves as the baseline for an activity report, which links a specific definition to a contributor (user) and a resource (creditor). All subsequent reward mechanisms are built upon these weighted definitions.

### Threshold-based achievement badges

Badges operate as cumulative, threshold-based algorithms designed to formalize and quantify sustained research effort by appealing to a contributor's intrinsic motivation and fostering consistency. They are defined deterministically by the partner resource based on an absolute volume of a specific set of activity types (e.g., "Submit 50 annotations"). The event-driven engine continuously evaluates the user's progress state against these configured thresholds, delivering the ability for the user to visualize progress. Furthermore, badges can be constrained by absolute date bounds, requiring thresholds to be met within a predefined time period. This enables platforms to incentivize bounded temporal engagement.

### Ranking-based competitive medals

In contrast to badges, which reward individual milestones, medals target extrinsic competitive drivers in end users through ranking-based achievements. Users compete for specific standings (e.g., 1st, 2nd, or nth place) based on criteria defined by the partner scientific resource. Like badges, medals support time constraints but are specifically configured to foster cohort-based competition. This ensures equitable comparison and maintains long-term engagement by allowing contributors to compete within appropriate tiers rather than against an insurmountable global leaderboard.

## System architecture

APICURON's architecture consists of three main components: a web server, a database, and a single-page application forming the user interface layer (Figure 6). Beneath this structure lies a modular, event-driven web server that underpins the recognition mechanism and enables efficient communication among system components. The persistence layer is built on MongoDB, a NoSQL document database chosen for its data structures, and it continues to serve as the primary data store. The core architectural improvement is the decoupling of event processing logic from the persistence layer. In this reactive architecture, event processing occurs within the server-side application layer, while MongoDB serves as the persistent repository for storing events and for displaying public rewards and contribution data. This separation preserves flexibility and allows the processing mechanism to be offloaded to dedicated infrastructure without modifying the database.

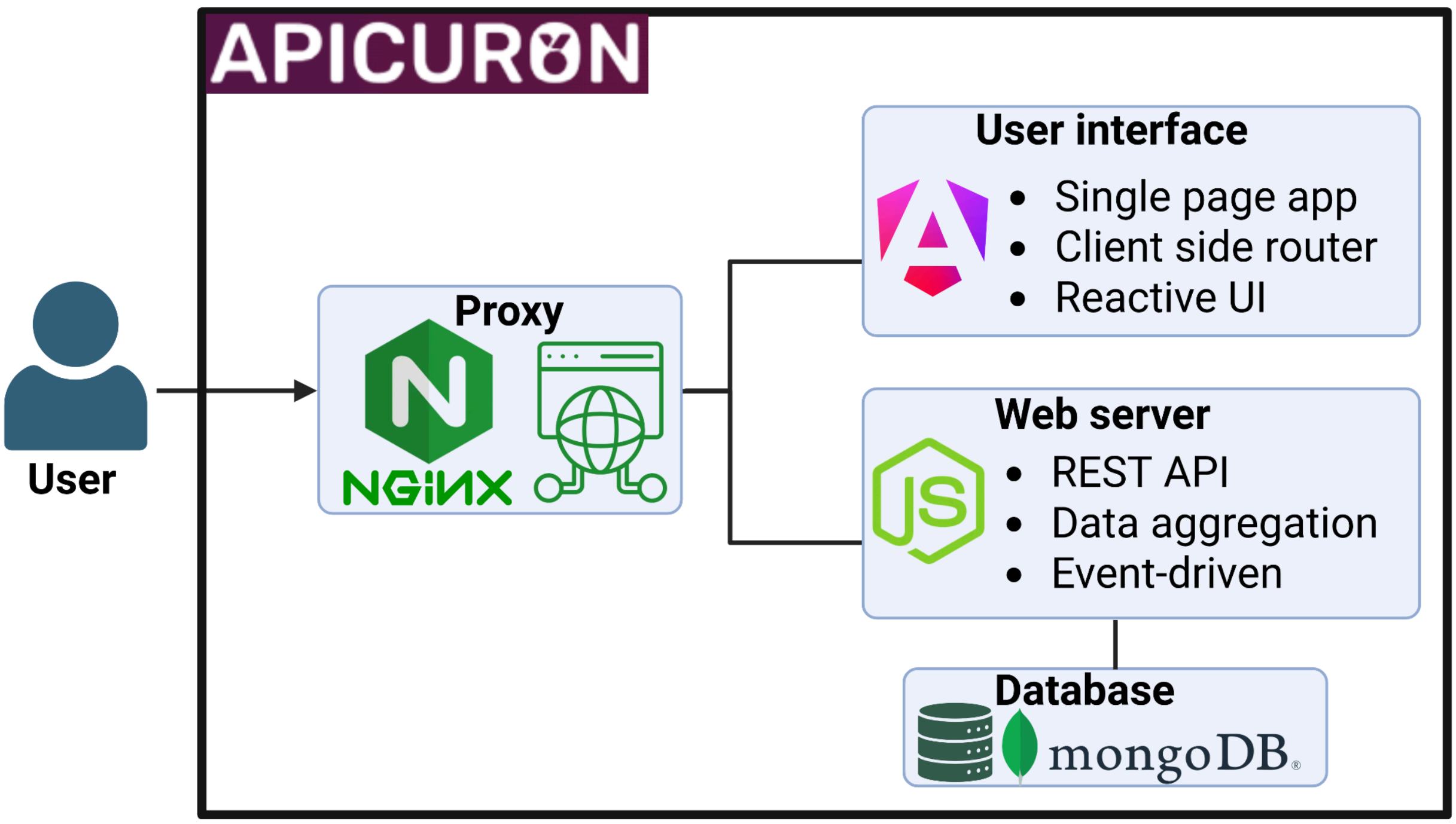

**Figure 6. APICURON system architecture.** The infrastructure utilises a decoupled, three-tier framework routed through an NGINX proxy. It comprises a reactive single-page application for the user interface, an event-driven Node.js web server handling the REST API and data aggregation, and a NoSQL MongoDB persistence layer. The core design isolates the asynchronous event-processing logic within the server-side application layer, separating it entirely from the database. This structural separation optimises non-blocking metric computation and allows the processing mechanism to be offloaded to dedicated infrastructure without altering the persistent data repository.

## Authentication, access control and ORCID integration

To ensure rigorous data provenance and accountability, APICURON enforces a strictly human-centric authentication model. Rather than utilizing generic service accounts or platform-centric credentials for partner resources, the system dictates that every operation modifying the database be explicitly traceable to a verified researcher. This identity verification is delegated entirely to ORCID via the OAuth 2.0 authorization framework. This integration serves a dual architectural purpose: it anchors system access to persistent, globally recognized academic identifiers, and it simultaneously secures the necessary permissions to propagate earned recognition directly to the biocurators' official ORCID records. At the infrastructure level, while the API exposes public endpoints to serve the user interface and general queries, all data-submission and sensitive routes are protected by strict role-based access control (RBAC).

## Data ingestion endpoints

APICURON utilizes a push-based integration model in which research infrastructures send contribution payloads via REST API calls supporting real-time submissions and bulk transfers. For timely submission, two HTTP POST endpoints are available. The /reports endpoint accepts an array of contribution events, enabling micro-batching of up to 50 reports, while a dedicated /reports/single endpoint optimizes high-frequency individual submissions and tolerates higher concurrency. For periodic or large-scale updates, the /reports/bulk endpoint accepts a JSON file containing a list of activity reports, accommodating platforms lacking event-driven systems and supporting periodic full log submission. The bulk endpoint supports redundancy reduction through global fields, allowing repeated fields within reports to be defined once and overridden when necessary. Processing of bulk submissions is deferred, and propagation of results depends on the dataset size. APICURON employs a Persistence-first ingestion strategy: upon receipt, valid contribution payloads are committed to the persistence layer immediately. Only after storage is successful is the HTTP acknowledgement returned, and the event is pushed to the in-memory processing queue, ensuring data durability. Once validated and ingested, payloads are handed off to the asynchronous processing tier to prevent high-volume submissions from blocking system resources.

## Partner Onboarding and Interoperability

A core challenge in establishing a centralized credit infrastructure for a fragmented research ecosystem is that different databases require entirely different definitions of units of work that are creditable. APICURON resolves this not by enforcing a universal contribution definition, but by providing a programmable recognition framework. Because a theoretical framework for credit attribution is only as valuable as its ability to integrate with existing infrastructure, a dedicated cohort of early adopters was engaged to stress-test the system's interoperability as APICURON transitioned from a prototype to a service. To sustainably scale the integration of new partner resources, a simple onboarding protocol has been introduced where prospective resources initiate integration through a dedicated registration pipeline. To facilitate metadata consistency and promote alignment with FAIR principles, this pipeline offers an optional integration with FAIRsharing registries. Prospective resources may leverage this feature to automatically retrieve and populate their institutional metadata, streamlining the onboarding process by minimizing manual data entry and reducing the potential for transcription errors. Following administrative vetting, approved resources are designated as official partners, and their managers receive the credentials required to interface with the core API.

To safeguard the integrity of the public APICURON version and ensure stringent compliance with data privacy standards, APICURON enforces an isolated staging phase prior to production deployment. A parallel development environment at [dev.apicuron.org](dev.apicuron.org) replicates the production architecture identically but is strictly federated with the ORCID Sandbox infrastructure ([sandbox.orcid.org](sandbox.orcid.org)). This architectural bifurcation dictates that partner resources must first validate their semantic mapping, event payload generation, and interoperability logic using synthetic testing identities. Only upon the successful algorithmic validation of their recognition schema in the staging environment are resources authorized to migrate to the public production instance. This enforces a strict boundary, isolating experimental integration logic from verified human biocuration data.

## Recognition engine and event processing

In APICURON, the primary unit of processing is a contribution event: a structured record linking a contributor, a partner resource, an activity type, a timestamp and optional contextual metadata such as a contribution group. Core logic resides in the Event-Driven Recognition Engine (Figure 7), responsible for continuous, non-blocking metric aggregation and reward computation. The outputs generated by APICURON are a function of two volatile, complex and interdependent states: the stream of contribution events and the evolving definitions of the Recognition Models. A mutation to either state introduces a cascade of updates to synchronize the outputs APICURON produces. Additionally, the range of these changes is inconsistent, ranging from continuous, singular updates to sporadic, massive dataset releases in which data providers refresh their full set of contribution events.

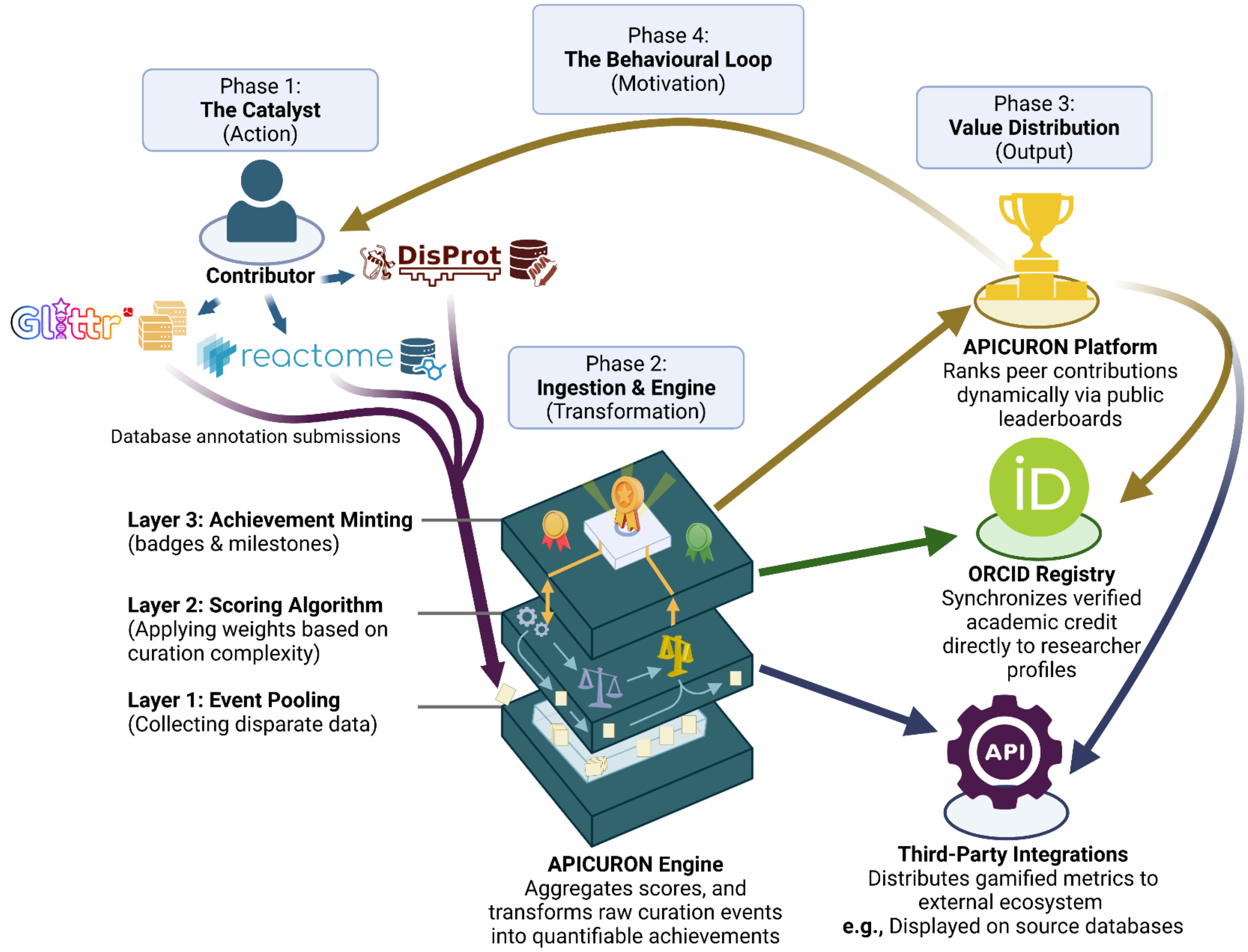

**Figure 7. The APICURON event-driven architecture and credit attribution cycle.** The platform captures contribution events from partner resources, processes them through the recognition engine and generates credit outputs, including scores, badges, medals and leaderboards. These outputs can be returned to partner resources through the public API or displayed directly on resource websites through a customizable leaderboard widget, which is already used by DisProt.

The initial conventional approach synchronizes such dynamic variables through scheduled batch processing, where the entire dataset is periodically re-evaluated to ensure consistency. Relying exclusively on this temporal approach introduces inefficiencies, thereby forcing the system to blindly recompute all aggregated data to capture even a minority of changes. The event-driven system resolves these inefficiencies by formalizing these changes as events. Every modification, whether a new contribution, recognition rule update, or full repository reload, is formalized as a discrete event and placed into a unified processing queue.

Processing is orchestrated via an internal task queue implemented within the Node.js runtime, leveraging asynchronous non-blocking execution in which event producers are isolated from the consumers. This abstraction provides two key advantages: First, it allows the processing logic to adapt dynamically to the type of event being handled, invoking specific handlers for contribution events, badge definition updates, or medal recalculations. Secondly, the in-memory queue functions as a base that enables future migration to a dedicated message broker for horizontal scaling without rewriting core processing logic.

Finally, the event abstraction supports crash recovery as well: if the in-memory queue is lost due to a service interruption, the system can leverage persisted raw reports to establish a consistent baseline. On startup, the system identifies unprocessed records and re-ingests them using the standard bulk event pipeline to recover a consistent database state. Consequently, the legacy requirement for batch processing effectively functions as a recovery strategy.

## Contribution group assignment model

Contribution Groups allow research platforms to define logical cohorts for isolated metric aggregation. To avoid the pitfalls of static user tagging, APICURON utilizes event-level contextualization, where group membership is captured at the moment of contribution. This approach tags the individual activity report rather than the user profile, to preserve historical integrity as researchers transition between roles. For example, if a community volunteer is subsequently hired as core staff, their indicator history accurately reflects this career trajectory without the data distortion inherent in static, global user tagging. This logic ensures that a single researcher can accumulate distinct sets of indicators and rewards corresponding to different phases of their career.

While the conceptual goal is to tag every contribution with its appropriate context, requiring that research platforms explicitly tag every individual API payload would impose an undue integration burden. To resolve this, APICURON implements a State-based Defaulting mechanism. Administrators manage cohort assignments through a management dashboard, establishing a "User-Group Binding" that acts as a forward-looking directive for future contributions. If a report lacks a specific group tag, the ingestion engine resolves the current binding and assigns the default group; in complex workflows, this can be overridden per report. Crucially, modifying a default group does not retroactively alter previous data, preserving the provenance of the scientific record.

# Data Availability

The datasets processed by APICURON during this study are publicly available via the APICURON REST API (https://apicuron.org/api/) and the APICURON web user interface. Certain subsets of the data are not publicly available because they contain personally identifiable information that could compromise the privacy of the APICURON users under the General Data Protection Regulation (GDPR). Access to this restricted data may be granted to qualified researchers upon reasonable request to the corresponding author, subject to a formal data sharing agreement. The aggregate data underlying Table 1 and Figure 2 are available through the public APICURON interface and API. Where individual-level records are restricted, equivalent aggregate summaries sufficient to reproduce the descriptive analyses reported here can be provided upon reasonable request.

# Code Availability

The underlying source code for APICURON's infrastructure is proprietary and maintained privately, but can be made available upon reasonable request. The platform and all features described in this manuscript remain freely accessible to the academic community as a software-as-a-service (SaaS) tool. Researchers can access the platform at apicuron.org. Documentation, user guides, and tutorials for interacting with the platform are available at apicuron.org/docs and apicuron.org/api/. To support reproducibility of the architectural description, public documentation includes API endpoint descriptions, payload schemas and example integration workflows. The recognition-model abstractions, event types and processing logic required to interpret the results are described in the Methods. Aggregate data used to generate the manuscript figures can be made available upon reasonable request, where they do not expose security-sensitive implementation details or personal data.

# Acknowledgements

The authors thank Patricia Palagi (Swiss Institute of Bioinformatics, Amphipôle, Quartier UNIL-Sorge, 1015 Lausanne, Switzerland) for the contributions across the Glittr.org resource integration and training artefact expertise provided during the APICURON project development.
The authors also thank the resource developers and APICURON profile holders for their contributions, feedback, and support during the development and evaluation of APICURON.

# Funding

This work has been supported by ELIXIR, the European infrastructure for life science data, through the ELIXIR Data Platform project (TECHNOLOGY-DATA-2024) [S.C.E.T., U.W.]. Additional funding from: European Union through NextGenerationEU PNRR project ELIXIRxNextGenIT (grant agreement no. IR0000010) [S.C.E.T.]; Horizon Europe projects EVERSE (grant agreement no. 101129744) [S.C.E.T.] and ELIXIR STEERS (grant agreement no. 101131096) [S.C.E.T., H.H.E]; Funded by the European Union. Horizon Europe MSCA Staff Exchange project IDPfun2 - grant agreement No. 101182949. [S.C.E.T.]. Views and opinions expressed are those of the author(s) only and do not necessarily reflect those of the European Union or the European Research Executive Agency. Neither the European Union nor the granting authority can be held responsible for them.